\documentclass[5p,twocolumn]{elsarticle}
\usepackage{graphicx,latexsym}
\usepackage{dcolumn}
\usepackage{amssymb,amsmath,bm}
\usepackage{subfigure}
\usepackage{braket}
\usepackage{siunitx}
\usepackage{array}
\usepackage{float}
\usepackage[nopar]{lipsum}
\usepackage{caption}
\usepackage{multirow}
\usepackage{bigstrut}
\usepackage{booktabs}
\newcommand{\ra}[1]{\renewcommand{\arraystretch}{#1}}
\usepackage[super]{nth}

\newcommand{\angstrom}{\text{\normalfont\AA}}
\usepackage{hyperref}
\hypersetup{
    pdfnewwindow=true,       
    colorlinks=true,         
    linkcolor=blue,          
    citecolor=blue,          
    filecolor=magenta,       
    urlcolor=black           
}

\usepackage[normalem]{ulem}

\def\sec#1{Sec.\ \ref{#1}}

\def\fig#1{Fig.\ \ref{#1}}
\def\tab#1{Tab.\ \ref{#1}}

\journal{}

\begin{document}

\begin{frontmatter}


\title{Conversion of the stacking orientation of bilayer graphene due to \break the  interaction of BN-dopants} 
	
\author[a1,a2]{Nzar Rauf Abdullah}
\ead{nzar.r.abdullah@gmail.com}
\address[a1]{Division of Computational Nanoscience, Physics Department, College of Science, 
             University of Sulaimani, Sulaimani 46001, Kurdistan Region, Iraq}
\address[a2]{Computer Engineering Department, College of Engineering, Komar University of Science and Technology, Sulaimani 46001, Kurdistan Region, Iraq}

\author[a1]{Hunar Omar Rashid} 

\author[a3]{Chi-Shung Tang}
\address[a3]{Department of Mechanical Engineering,
	National United University, 1, Lienda, Miaoli 36003, Taiwan}

\author[a4]{Andrei Manolescu}
\address[a4]{Reykjavik University, Department of  Engineering,
	Menntavegur 1, IS-102 Reykjavik, Iceland}     

\author[a5]{Vidar Gudmundsson}   
\address[a5]{Science Institute, University of Iceland,
	Dunhaga 3, IS-107 Reykjavik, Iceland}


\begin{abstract}
	
A conversion of AA-  to AB-stacking bilayer graphene (BLG) due to interlayer interaction is demonstrated. Two types of interlayer interactions, an attractive and a repulsive, between the Boron and Nitrogen dopant atoms in BLG are found. 
In the presence of the attractive interaction, an AA-stacking of BN-codoped BLG is formed with a less stable structure leading to weak mechanical properties of the system. Low values of the Young modulus, the ultimate strength and stress, and the fracture strength are observed comparing to 
a pure BLG. In addition, the attractive interaction induces a small bandgap that deteriorates the thermal 
and optical properties of the system. 
In contrast, in the presence of a repulsive interaction between the B and N atoms, the AA-stacking is converted to a AB-stacking with a more stable structure. Improved mechanical properties such as higher Young modulus, the ultimate strength and stress, fracture strength are obtained comparing to the AA-stacked BN-codoped BLG. Furthermore, a larger bandgap of the AB-stacked bilayer enhances the thermal and the optical characteristics of the system.

\end{abstract}


\end{frontmatter}

\section{Introduction} 

Graphene has been considered as a promising material for applications in the fields of sensors, photonics, and electronic devices because of its excellent characteristics such as optical transparency, low density, high carrier mobility, and chemical stability \cite{KwongHongTsang2019, Chen2019, McCann_2013, gudmundsson2019coexisting}. Combining two layers of graphene in a specific configuration called bilayer graphene is also
a high-potential material with possible applications in electronics and optics \cite{PhysRevB.78.045405, PhysRevB.75.155430, TANG20171529, ABDULLAH2020114221}. 
It has been experimentally shown that BLG has a number of remarkable characteristics such as  outstanding electrical, mechanical and chemical performance leading to a great flexible transparent electrodes used in touch-screen devices \cite{Bae2010},
high-switching ratio digital transistors \cite{doi:10.1021/nl104000b}, and efficient infrared detectors \cite{Zhang2009}. 
Stiffness and flexibility properties of BLG make it a great candidate for fuel cells and a material for use in structural composite applications \cite{Dikin2007}.

The aforementioned properties of BLG can be improved by tuning the interaction between the graphene layers, or the van der Waals interactions appearing between the layers \cite{PhysRevB.74.205434, TANAKA1997121}. The DFT-D technique, dispersion-corrected density functional theory, has been used to describe the interlayer interaction energy of BLG with high accuracy \cite{C0CP02614J}. 
It has been found that the interlayer electron motion is affected by the atomic orientation of the two layers \cite{Alden11256, Yan2013}. The atomic orientation controls the strength of the interlayer van der Waals bonding.
The interlayer electron interaction is influenced by the stacking configuration of the BLG 
leading to interesting physical properties.
For instance, a moiré pattern has been observed due to the interaction between the layers for twisted BLG \cite{PhysRevLett.120.156405} with extraordinary optical properties found \cite{doi:10.1063/1.4776694, ABDULLAH2019102686, doi:10.1021/jp504222m}. Furthermore, a relatively weak strength of interlayer interaction, and low energetic cost of relative translation of the layers, has been seen for AA-stacking BLG \cite{Vuong_2017}.
Compared with other stacking structures, AA- and twisted-stacking, very strong coupling exists between the two layers of AB-stacked BLG, 
which has the lowest energy and the most stable structure among the different orientations \cite{Chen2013}. Strong coupling between the layers improves and leads to numerous physically interesting properties in BLG.  
In recent experimental work, the stacking orientation of BLG is investigated and it is found that the orientation of BLG could be changed from weak stacking coupling to AB strong stacking coupling. In addition, a reduction in the interlayer distance is considered in a high-pressure environment to detect orientation changes \cite{CHEN2021480}.  We therefore, believe that the stacking configuration of BLG is still challenging and further work is needed to clearly the situation. 

In the present paper, we try to control the distance and interlayer interactions of AA-stacked BLG by boron (B) and Nitrogen (N) dopant atoms. We will show how the AA-stacking with a weak interlayer interaction can be converted into a AB-stacking with a strong interlayer interaction by doping of B and N atoms in the BLG. More precisely, the conversion of AA- to AB-stacked BLG will be achieved by controlling the dimer positions of the B and N atoms in the hexagonal structure of the graphene \cite{ABDULLAH2020103282}. In addition, the electronic, mechanical, thermal, and optical properties of the system will be shown for both AA- and AB-stacked BLG, and one can see significant improvement of the physical properties of the AB-stacked BLG in our study~\cite{Abdullah2019, Abdullah_2019}.

In \sec{Sec:Model} the BLG structure is briefly over-viewed. In \sec{Sec:Results} the main achieved results are analyzed. In \sec{Sec:Conclusion} the conclusion of the results is presented.

\section{ Computational technique}\label{Sec:Model}

We assume a pure BLG and BN-doped BLG with different B and N atom configurations. 
The density functional theory (DFT) techniques based on the local density approximation (LDA)  implemented in Quantum espresso package have been used to study physical properties of our models \cite{Giannozzi_2009, giannozzi2017advanced}.
A periodic boundary conditions 
are applied to a $5.65 \, \angstrom \times 2.44 \, \angstrom \times 20 \, \angstrom$ model cell of pure BLG. 
The Brillouin zone is sampled with $k$-point grids from $10 \times 10 \times 1$ to  $20 \times 20 \times 1$, 
and the basis set is plane waves with a maximum kinetic energy of $680\text{-}1360$~eV \cite{RASHID2019102625}. 
The spacing of the real space grid used to calculate the Hartree,
the exchange and the correlation contribution of the total energy $1360$~eV.
The van der Waals interaction is included in the exchange (XC) functional, and
the structures are relaxed until the forces on each atom were less than 0.001 eV/$\angstrom$.

The XCrySDen is used to visualize all the structures \cite{KOKALJ1999176, ABDULLAH20181432}. In addition, the Boltzmann transport
properties software package (BoltzTraP) is employed to study the thermal properies of the systems \cite{madsen2006boltztrap-2}. The BoltzTraP code uses a mesh of band energies and has an interface to the QE package. The optical characteristics of
the systems are obtained by the QE code. Optical calculations are performed using 
a $100 \times 100 \times  2$ optical mesh and an optical broadening of $0.1$~eV.

\section{Results}\label{Sec:Results}

\begin{figure}[htb]
	\centering
	\includegraphics[width=0.45\textwidth]{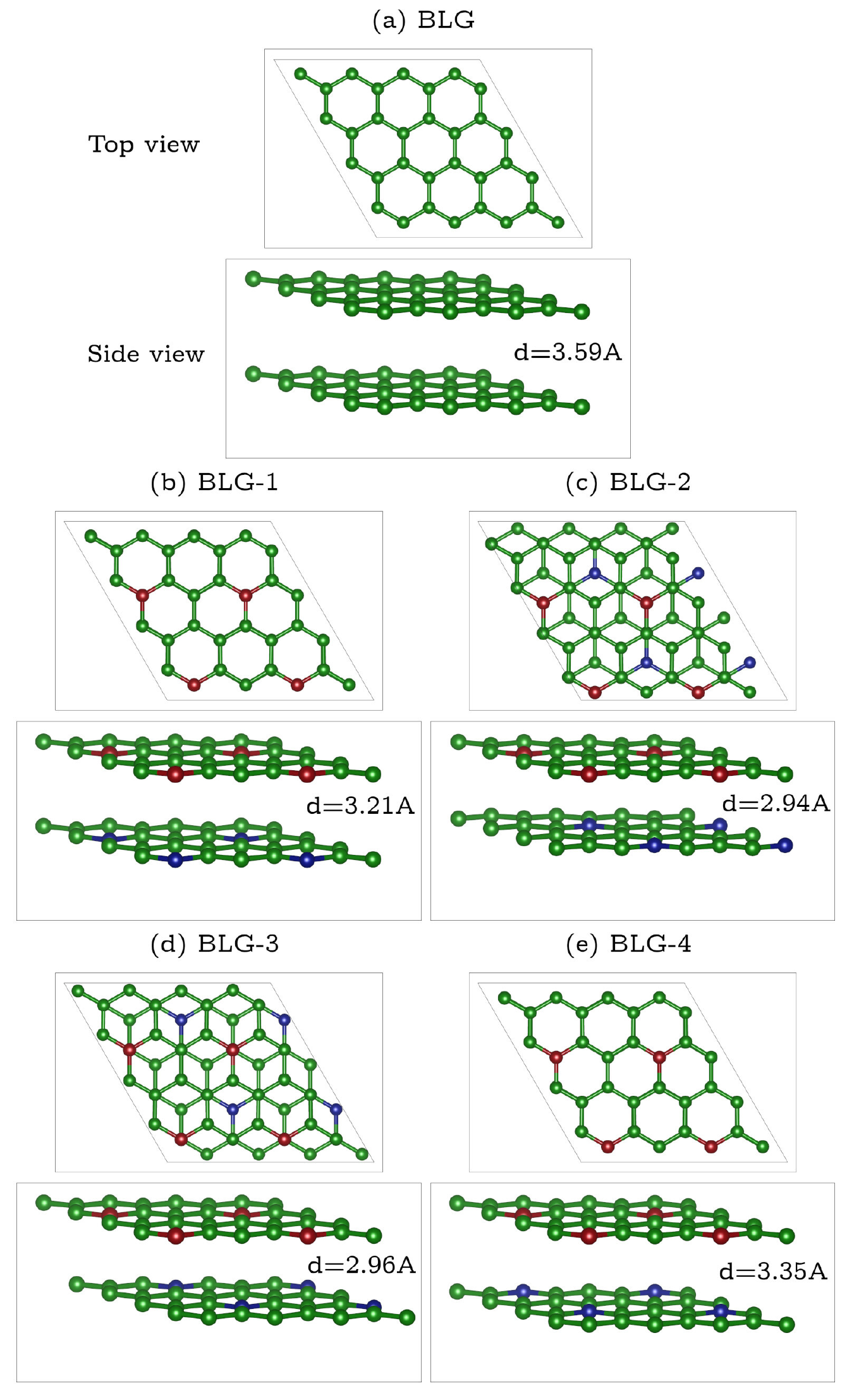}
	\caption{Pristine BLG (a) and BN-codoped BLG (b-d)  where the C, B, and N atoms are green, red, and blue colored, respectively, the B(N) atom is doped in top(bottom) layer.  The B atoms is fixed at a para-position in the top layers of all structures, but the 
		N atom is doped in a para-position at a site corresponding to the B atom in BLG-1, a meta-position in BLG-2, ortho-position in BLG-3, and para-position at a site opposite to the B atom in BLG-4 in the botthom layer. $d$ indicates the interlayer distances.}
	\label{fig01}
\end{figure}

Pristine BLG and BN-codoped BLG are shown in \fig{fig01}, with the B (red) atom fixed at a para-position in the top layer, and the N (blue) atom is doped in the bottom layer at a para-position corresponding to the B atom (BLG-1), the meta-position (BLG-2), the ortho-position (BLG-3), and a para-position at a site opposite to the B atom (BLG-4). 
The position of the B and N atoms together forms different isomers which are called a para-isomer for BLG-1, a meta-isomer for BLG-2, an ortho-isomer for BLG-3, and a para-isomer for BLG-4.

The structural parameters calculated for BLG and BN-codoped BLG are presented in \tab{table_one}. 
The calculation of the structural parameters indicate that in the presence of the B and N atoms 
the average bond lengths of C-C, C-B, and C-N are slightly changed for all BN-codoped BLGs. 
This is due to the larger and the smaller atomic radius of the B and the N atoms, respectively, compared to the atomic radii of a C atom \cite{ABDULLAH2020126807}. 
Furthermore, the average lattice parameters of the BN-codoped structures are increased compared to the pure BLG indicating a super-cell expansion of the BN-codoped BLG.
\begin{table*}[h]
	\centering
	\begin{center}
		\caption{\label{table_one} The average bond lengths of C-C, C-B, and C-N, lattice parameters, $a$, interlayer distance, $d$, interlayer interaction energy, E$_{\rm in}$, and distance between B and N atoms, $d_{\rm BN}$,  for BLG and BN-codoped BLG structures. The unit of all parameters except  $\Delta E$ is $\angstrom$.}
		\begin{tabular}{|l|l|l|l|l|l|l|l|}\hline
			Structure	 & C-C     & C-B       & C-N          &   a       &  d         & $\Delta E$  (eV)  & d$_{\rm BN}$\\ \hline
			BLG	              & 1.42    &   -----    &    -----         & 2.44   & 3.59    & -----                         & ---         \\
			BLG-1	        & 1.413  &  1.46    &   1.417     & 2.468  & 3.21    &  -3.25                     & 3.21        \\
			BLG-2	        & 1.418  &  1.458  &   1.419     & 2.471  & 2.94    &   2.45                     & 4.11        \\
			BLG-3	        & 1.415  &  1.458  &   1.419     & 2.471  &  2.96   &   2.39                      & 4.1          \\   
			BLG-4	        & 1.416  &  1.458  &   1.418     & 2.465  &  3.35   &  -4.35                     & 4.4            \\   \hline
	\end{tabular}	\end{center}
\end{table*}
It is found that the C-C bond length, the lattice parameters, and the interlayer spacing of our pure BLG are consistent with previous studies \cite{RevModPhys.81.109,  ABDULLAH2020126350}. The C-B and C-N bond lengths and lattice parameters
of BN-codoped BLG agree very well with Alattas and Schwingenschl{\"o}gl \cite{Alattas2018}.

\lipsum[0]

\subsection{Interlayer Interaction}

The key point of this work is the interlayer interaction due to the B and N doped atoms.
There are several details that should be considered to analyse the components or the sources of the interlayer interactions in BLG.
We take into account the van der Waals interactions which introduce a non-bonding potential between the layers of BLG.
The Van der Waals interaction is a dipole-dipole interaction that can dominate other 
interactions if the distance between dipoles is around and beyound $4\text{-}5$~$\angstrom$~\cite{PhysRevB.80.245424, PhysRevB.62.13104, ABDULLAH2020113996}.  
This type of interaction does not play a key role in our systems 
as the distance between the layer does not reach that limit (see \tab{table_one}).
Other types of interactions dominate over the weak dipole-dipole interaction
in our system. More important is the interaction that arises due to the sp$^3$ bonding between 
the two layers of the BLG. This type of interaction is best studied by incrementally moving two 
atoms with the same planar coordinates (one in the upper layer and the other
one in the lower layer) to the sp$^3$ bond distance which is about $1.54$~$\angstrom$ 
and allow the structure to relax \cite{doi:10.1063/1.4740259}. Again, we do not expect 
this type of interaction to be important as our models have larger interlayer distance.
The third type of interaction is a nonbonding interaction between the layers of a BLG 
that could be either a repulsive or attractive \cite{doi:10.1021/jp2095032}. 
This type of interaction is effective if the interlayer distance 
is around $2\text{-}4$~$\angstrom$. We assume this type of interaction to play an important 
role between the layers, and the interaction energy between two dopant atoms in a structure 
is given by \cite{doi:10.1063/1.4742063}
\begin{equation}
	\Delta E = E_2 - E_0 + 2 \times E_1 
\end{equation}
where E$_0$, E$_1$ and E$_2$ are the total energies of the systems with zero, one, and two dopant atoms, respectively.  The obtained interaction energies of all BN-codoped systems are presented in \tab{table_one}.
It has been shown that the interaction energy in BN-codoped graphene varies inversely with the
distance between the B and N atoms because of the nature of the Coulombic electrostatic interaction,
and the interaction strength is almost zero when the separation distance is greater than or equal to $4.0$~$\angstrom$  \cite{doi:10.1063/1.4742063}.  
The value of $\Delta E$ can have a negative or a positive sign indicating that the interaction 
between the B and N atoms is attractive or repulsive, respectively.

At first the sight, one can see the interaction between the B and the N atoms leading to an attractive between the layers of BLG-1, as the interaction energy is negative when both the B and N atoms sit on the same isomer position, para-position, but in different layers. Consequently, the 'Coulomb electrostatic potential' between the B and N atoms is directly along the $\pi$-bond directions on the same line perpendicular to both the B and N atoms.  The Coulomb electrostatic force only affects the interlayer spacing in the $z$-direction and does not influence the atomic positions of both layers in the $xy$-plane.  In this case, the AA-stacking behavior of BLG-1 is unchanged. 
The same scenario can be applied to BLG-4 when the B and N atoms sit in opposite para-positions. The separation between the B and N atom is large here, $4.4$~$\angstrom$,  leading to a weak interaction energy and interlayer interaction. So, the interaction between the B and N atom does not influence the structural property of the system and the AA-stacking of  BLG-4 is again unchanged. Notice that the interlayer distance of BLG-4 is maximum among all BN-codoped structures and it is also close to the interlayer distance of pure BLG confirming a weak interaction between the dopant atoms.

The most interesting point here is the interlayer interaction in BLG-2 and BLG-3 due to the relative position of the B and N atoms that does not only decrease the interlayer distance, but also converts the AA-stacked BLG to an AB-stacked one  (see \fig{fig01}c-d). This is the key point of our study. 
In these two structures, the B and N atoms are doped in such a way that a meta-isomer in BLG-2, and an ortho-isomer in BLG-3 are formed as we mentioned before. In these two isomers, the 'tangential Coulomb electrostatic force' present between the B and N atoms generates a repulsive force 
between the B and N atoms. 
\lipsum[0]
\begin{figure*}[htb]
	\centering
	\includegraphics[width=0.9\textwidth]{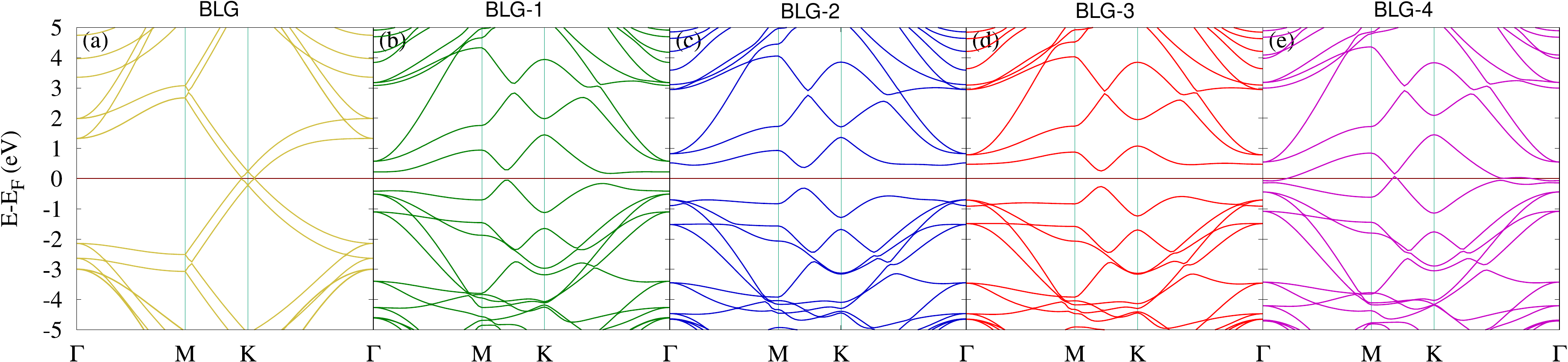}
	\caption{Dispersion energy or electronic band structure of pure BLG (a), and BN-codoped BLG (b-e) are shown. The Fermi energy is set to zero.}
	\label{fig02}
\end{figure*}
Consequently, a diffusion and depletion between the layers emerges 
activating a convertion from the AA-stacking to an AB-stacked structure.  Our results for the interaction energy shown in \tab{table_one} confirms the repulsive force between the B and N atoms in these two structures as their interaction energies are positive. The diffusion process between the layers continues until the distance between the B and N atoms approaches $4.0$~$\angstrom$, which is the equilibrium imposed by the effective interaction of the two dopant atoms forming an effective interaction \cite{doi:10.1063/1.4742063}. The distance between the B and N atoms after full relaxation in BLG-2 (BLG-3) is $4.11$ ($4.1~\angstrom$), respectively.  Any change of physical properties including 
the layer conversion arising from the isomer position of the dopant atoms is called the effect of isomerization~\cite{RANI201428, ABDULLAH2020114556}.

\subsection{Band structure and DOS}

In order to understand the physical properties of BN-codoped BLG, such as thermal and optical characteristics, we need to present the dispersion energy of the structures. In \fig{fig02}, the dispersion energy or the electronic band structures of pure BLG (a), and BN-codoped BLG (b-e) are 
displayed.
In \fig{fig02}(a) the dispersion of pure BLG has multiple linear dispersion bands forming $\pi_1$,  
$\pi_2$ and $\pi_1^*$, $\pi_2^*$ that touch at the K-point of the first Brillouin zone. So, the intersection of the conduction and valence bands is located symmetrically near the K-point. The linear dispersion bands are mainly due to the electronic interlayer coupling that is suppressed by the Pauli repulsion between the graphene layers \cite{OuldNE2017}. The linear dispersion of the AA-stacked BLG has been experimentally confirmed \cite{Kim2013}.

The computed electronic band structures of BLG-1, BLG-2, and BLG-3, based on LDA calculation, show  a semiconducting property with an indirect bandgap of BLG-1, and a direct bandgap of the BLG-2 and BLG-3 structures. The band structure of BLG-4 exhibits a metalic behavior as the lowest conduction band crosses the Fermi energy near the $\Gamma$-point. 
The origin of bandgap opening comes from the redistribution of surface charge due to the B and N atoms that breaks the local symmetry of the BLG.  The redistribution of the charge density of the BN-codoped BLG gives rise to the separation of the valence band and conduction band at the Dirac point.  The B atoms lead to band shift below the Fermi level, and the N atoms above
the Fermi level, such that the counterbalance of both creates a gap at the Fermi level. 
Furthermore, the electrostatic Coulomb dipole interaction between the B and N atoms is also important in breaking the symmetry of BLG and thus to the band gap opening \cite{RANI201428}. A narrowing of the bandgap in BLG is also induced by interlayer interaction~\cite{doi:10.1021/nl802409q}. The computed bandgaps of the BLG-1, BLG-2, and BLG-3 are $0.346$. $0.685$, and $0.515$~eV, respectivily. It is interesting to see that the bandgaps of the BLG-2 and BLG-3 with the repulsive interaction between the B and N atoms are relatively larger than those of the BLG-1 and BLG-4 with the attractive interaction between the B and N atoms.
It has been confirmed that the repulsive interaction between the layers forms a direct and large bandgap, while an attractive interaction 
induces an indirect and small bandgap in both the AA- and the AB-stacked BLG with B and N dopant 
atoms \cite{ABDULLAH2020100740}.
Our results for the band dispersion and the bandgaps agree very well with recent results for BN-codoped BLG in which the BLG is doped with one B atom in one layer and one N atom in the other \cite{Alattas2018}.

\begin{figure}[htb]
	\centering
	\includegraphics[width=0.48\textwidth]{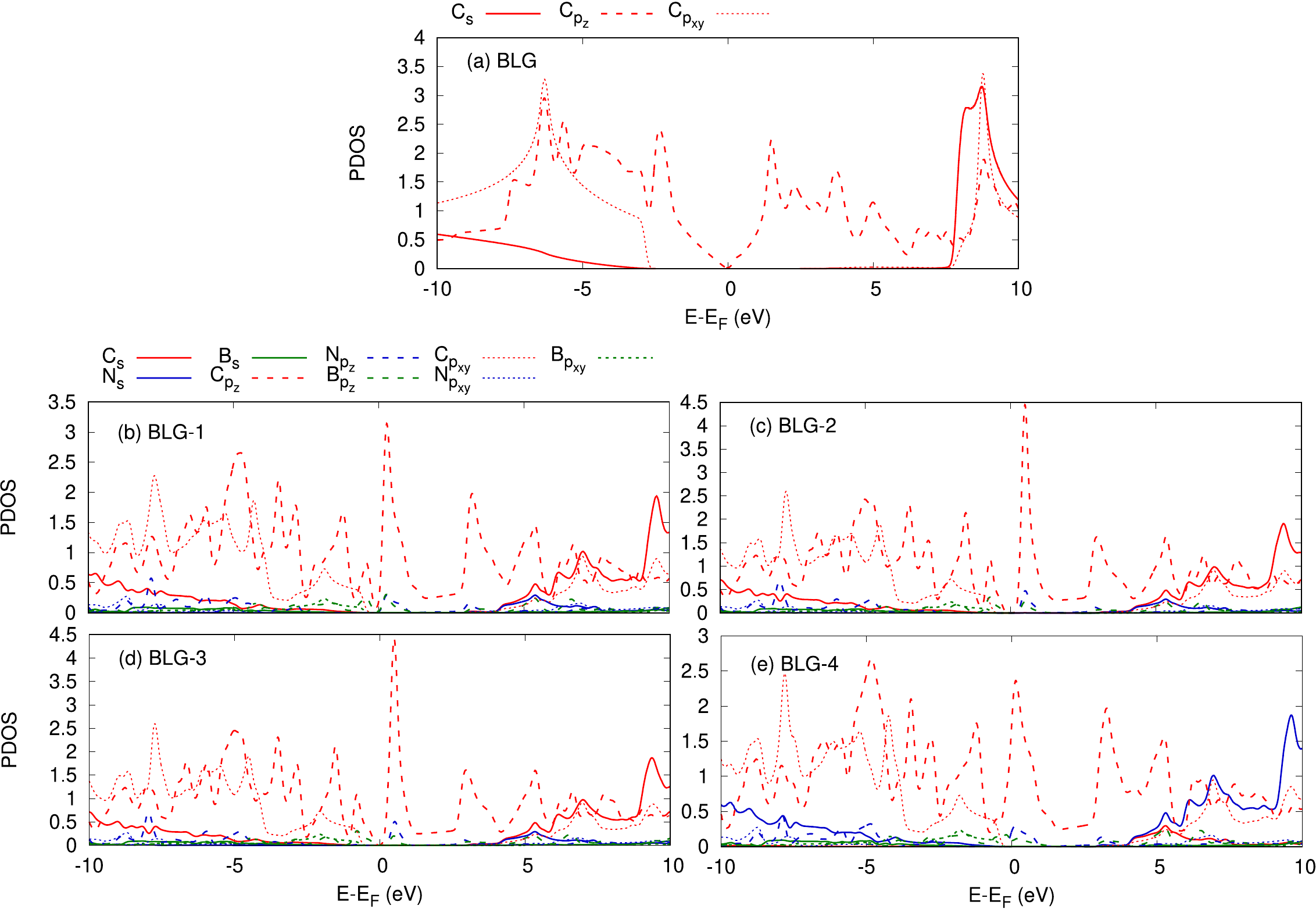}
	\caption{Partial density of states (PDOS) of pure BLG (a), and BN-codoped BLG (b-e) are shown. 
	         The Fermi energy is set to zero.}
	\label{fig03}
\end{figure}

The computed partial density of states (PDOS) of pure BLG and BN-codoped BLG are presented in \fig{fig03}.  It is clearly seen that the maximum of the valence band and the minimum of the conduction band for pure-BLG are totally formed by the four $p_z$ $\pi$-bands  
around the Fermi energy.
Furthermore, the $s$ and $p_{x,y}$ contribute to the lower part of the valence band and upper part of the conduction band as is expected for pure BLG~\cite{MOHAN20121670, abdullah2019thermoelectric}.

\begin{table*}\centering 
	\ra{1.3}
	\begin{tabular}{@{}rrrrcrrrcrrr@{}}\toprule
		& \multicolumn{4}{c}{Zigzag} & \phantom{abc}& \multicolumn{4}{c}{Armchair} \\
		\cmidrule{2-5} \cmidrule{7-10} 
		& YM (TPa)         & UTS        & F-strength         &  F-strain ($\%$)     && YM (TPa)  & UTS    & F-strength        &  F-strain ($\%$)    \\ \midrule
		BLG         & 0.99                  & 99.47     & 99.47    &  15.12                        && 0.99          & 96.06 & 96.06   &   12.66                      \\
		BLG-1      & 0.65                 & 58.55     & 58.55    &   12.14                       && 0.65          & 56.73 & 56.73   &    11.24                     \\
		BLG-2      & 0.856               & 74.89     & 70.58    &   12.45                       && 0.858       & 77.66 & 61.73    &    18.07                    \\
		BLG-3     & 0.855               & 77.82      & 77.82    &   12.14                       && 0.875       &78.04  & 71.58    &    13.76                    \\
		BLG-4     & 0.608               & 55.67      & 55.67    &   12.14                       && 0.622       & 59.92 & 59.92     &   11.24                    \\
		\bottomrule
	\end{tabular}
	\caption{The Young modulus (YM), the ultimate strength (UTS), the fracture strength (F-strength), and the fracture strain (F-strain)  for BLG and BN-codoped BLG structures in the zigzag and armchair directions. The unit of UTS and F-stress is GPa.}
	\label{table_two}
\end{table*}

The PDOS of the BN-codoped BLG is analyzed showing that in all four cases of a BN-codoped BLG the higher valence bands have a contribution from the $p_z$ of the C and B  atoms, and the lower conduction bands are formed by the $p_z$ of the C, B and N atoms. 
We note that the major contribution to the $\pi$-bands comes from the C atoms.
More precisely,  the contribution of the B and N atoms around the Fermi energy is slightly 
higher for the BLG-2 and BLG-3, where the a repulsive interaction between the B and N atoms exists.  
It is also expected that the contribution of both the $s$ and $p_{x,y}$ by the B and N atoms are 
increased at the lower valence and the upper conduction bands.

\subsection{Mechanical properties}

The DFT calculations can be used to investigate the stress-strain curves of our systems. 
We apply uniaxial tensile simulations to probe the stress-strain properties. 
The load is gradually applied to a specific direction of the structure,  the zigzag or armchair directions~\cite{MORTAZAVI2017228}. During a uniaxial tensile loading, the periodic dimension along the loading direction is increased step-by-step with a fixed strain of $0.02$.

\begin{figure}[htb]
	\centering
	\includegraphics[width=0.35\textwidth]{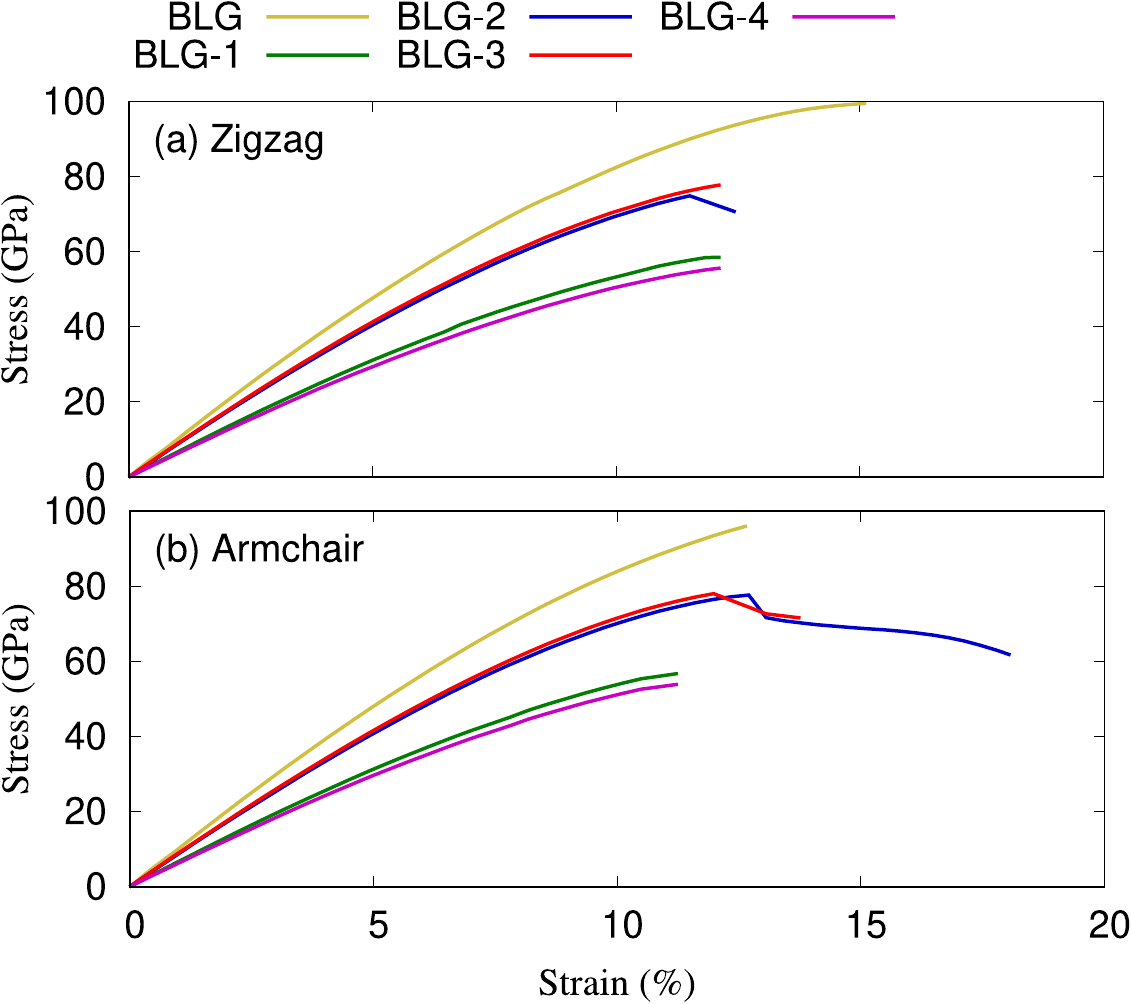}
	\caption{Stress-strain curves for pure BLG (golden),  BLG-1 (green), BLG-2  (blue), BLG-3 (red), and BLG-4 (purple) in the zigzag (a) and armchair (b) directions.}
	\label{fig04}
\end{figure}

The stress-strain curves of pristine BLG as well as the ones for the BN-codoped BLG are presented in \fig{fig04} for zigzag (a), and armchair (b) directions. The strain-stress curves depend on stability of the structures in which the most stable structure has the highest 
stress-strain curve. One can see that the stress-strain curves of BN-codoped BLG are lower than those for the pure BLG for both directions over the entire values of the strain. This is due to the more stable C–C bond than the B–C or N–C bond. The bond energy of C-C is $3.71$~eV which is greater than that of C-N ($2.83$~eV) and C-B ($2.59$~eV) \cite{NOZAKI199641, Bhandary2012GrapheneBoronNC}. 
Consequently, the structure with a higher number of C–C bonds will be more stable than the others. 
This is the reason for higher stress-strain curves of pure the BLG compared to the BN-codoped BLG.

In addition, a structure with the lowest total energy will be the most stable structure \cite{OuldNE2017}. Our DFT calculations of the total energy confirm that the most stable structures among the BN-codoped BLG are BLG-2 and BLG-3. These two structures have lower total energies than 
BLG-1 and BLG-4. Therefore, the BLG-2 and BLG-3 structures with repulsive interaction between the B and N atoms have higher stress-strain curves than the BLG-1 and BLG-4 structures with an attractive interaction.  This also refers to AB-stacked BLG-2 and BLG-3 formed by the repulsive interaction in which the AB-stacked shapes have more energetically stable structure than AA-stacked one.

The values of the ultimate strength (UTS), the Young modulus (YM), the fracture strength (F-strength), and the fracture strain (F-strain) can be computed and their values for all structures are presented in \tab{table_two}.
 It is known that the YM defines the tensile stiffness of a structure. 
In a small range of strain of about a few percent (strain$<\%4$),  when the tensile strain increases, the stress that the system experienced enlarges linearly for the BLG and BN-codoped BLG structures.  From the linear regime of the curves, one can calculate the Young’s moduli, which are basically the initial slope of the stress–strain curves. The strain-stress curve of pure BLG reveals the YM to be $0.99$~TPa for both the zigzag and armchair directions in good agreement with the theoretical value of $0.96$~TPa \cite{liu2018molecular}, and  the experimental value of $1.0$~TPa \cite{Lee385-2}.

When a large strain is applied, the stress response of a system is non-linearly up to the strain 
for which the system fails.
The UTS defines the maximum stress that a structure can withstand while being stretched. The UTS for pure BLG in the zigzag direction is slightly higher than that of the armchair one. Furthermore, the F-strength, also known as breaking strength, is the stress at which a structure fails via fracture, while F-strain reflects flexibility of the system. The values of the F-strength and the F-strain in the zigzag direction of the pure BLG is higher than that of the armchair direction which can be related to the anisotropy of the system.

It is interesting to see that the YM, UTS, F-strength, and F-strain of the BLG-1 and BLG-4 are all decreased compared to the pure BLG. 
These two structures have the AA-stacking shape and they are less stable leading to smaller 
values of the YM, UTS, F-strength, and F-strain. 
In contrast, the BLG-2 and BLG-3 with the AB-stacking shape are more stable leading to 
higher YM, UTS, F-strength, and F-strain compared to the 
BLG-1 and BLG-4 structures. We should remember that all the BN-codoped BLG structures have the same number of C-B and C-N bonds, we therefore emphasize that the number of C-B and C-N bonds does not 
play a role when we compare the stress-strain curves of these four structures.

\subsection{Thermal properties}

The thermal characteristics of our structures at the low temperature ranging from $20$ to $160$ K are considered where the phonons are inactive \cite{PhysRevB.87.241411, ABDULLAH20181432}.

\begin{figure}[htb]
	\centering
	\includegraphics[width=0.35\textwidth]{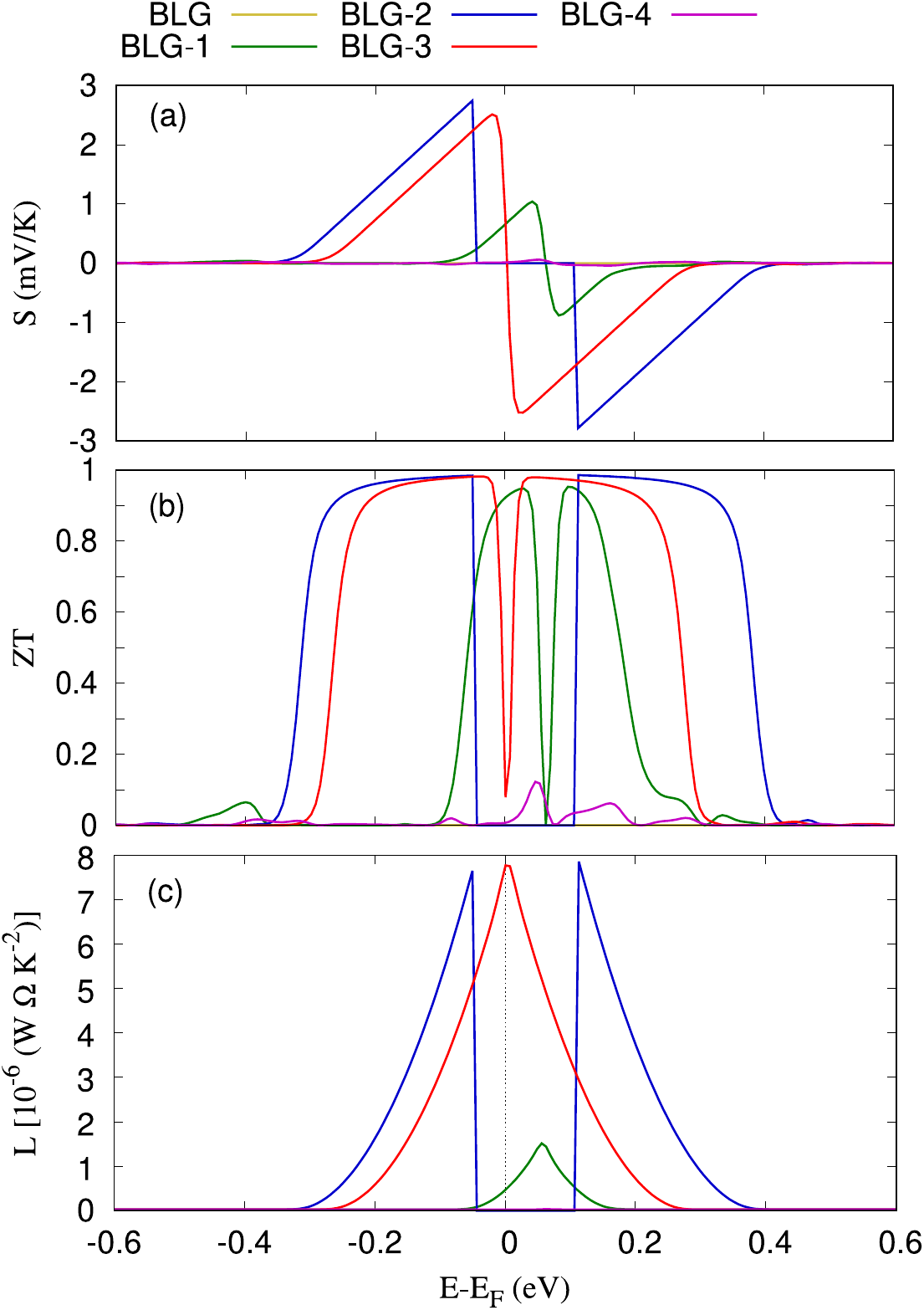}\\
	\caption{Seebeck coefficient (a), figure of merit ($ZT$), and Lorentz number ($L$) versus energy for pure BLG (golden),  BLG-1 (green), BLG-2  (blue), BLG-3 (red), and BLG-4 (purple). The Fermi energy is set to zero. }
	\label{fig05}
\end{figure}

In such a low temperature range, the electrons deliver the main contribution to the thermal
properties. Figure \ref{fig05} presents the Seebeck coefficient, $S$, (a), the figure of merit, 
$ZT$, (b), and Lorentz number, $L$, (c) as a function of energy for pure BLG and BN-codoped BLG structures.
A good thermoelectric material should have a a low thermal conductivity, and a high $S$ and electrical conductivity. The thermoelectric performance of a monolayer and a BLG is poor because of closed bandgaps, leading to a small $S$ \cite{Sadeghi2015, ABDULLAH2020126578}. 
This is also seen for pure BLG shown in \fig{fig05}(a), where the $S$ has an extremely small value which is almost invisible in the figure. 
Similarly, this gives rise to very small value of $ZT$ and $L$ for pure BLG as well.

Asymmetric peaks in the PDOS close to the highest occupied
state shown in \fig{fig03}, and the opening up of a bandgap are expected to increase the $S$, $ZT$ 
and $L$ in BLG-1, BLG-2 and BLG-3 structures \cite{pilemalm2019effect,abdullah2019photon}. 
A small bandgap of BLG-1 induces small $S$, $ZT$ and $L$, while the largest bandgap of BLG-2 gives the maximum values of $S$, $ZT$ and $L$. Therefore, one can expect a higher thermoelectric performance for the BLG-2.

\subsection{Optical responses}
 
We explore the optical properties of pure BLG and BN-codoped BLG which may be interesting for optoelectronic use of graphene-based materials. The rise in interest of BLG in optoelectronics is presented by its applications ranging from solar cells and light-emitting devices to touch screens, photo-detectors and ultrafast lasers. These all come from the unique optical and electronic
properties of BLG and their relevance for nano-photonics \cite{Bonaccorso2010, Abdullah_2019, Huang2019, ABDULLAH2018223}. 
  
\begin{figure}[htb]
	\centering
	\includegraphics[width=0.4\textwidth]{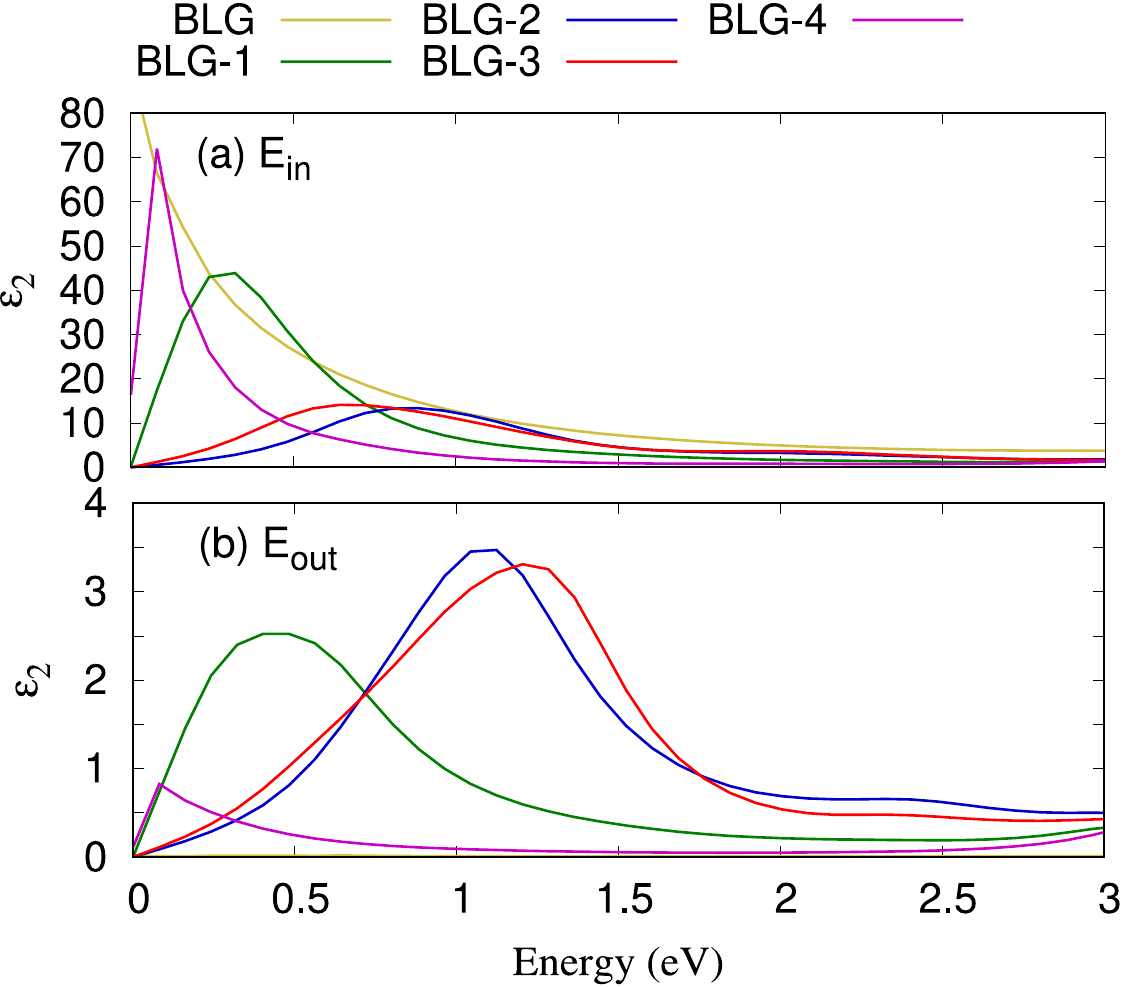}
	\caption{Imaginary part of dielectric function, $\varepsilon_2$, versus energy for pure BLG (golden),  BLG-1 (green), BLG-2  (blue), BLG-3 (red), and BLG-4 (purple), in the E$_{\rm in}$ (a), and  E$_{\rm out}$ (b).}
	\label{fig06}
\end{figure}

\begin{figure}[htb]
	\centering
	\includegraphics[width=0.4\textwidth]{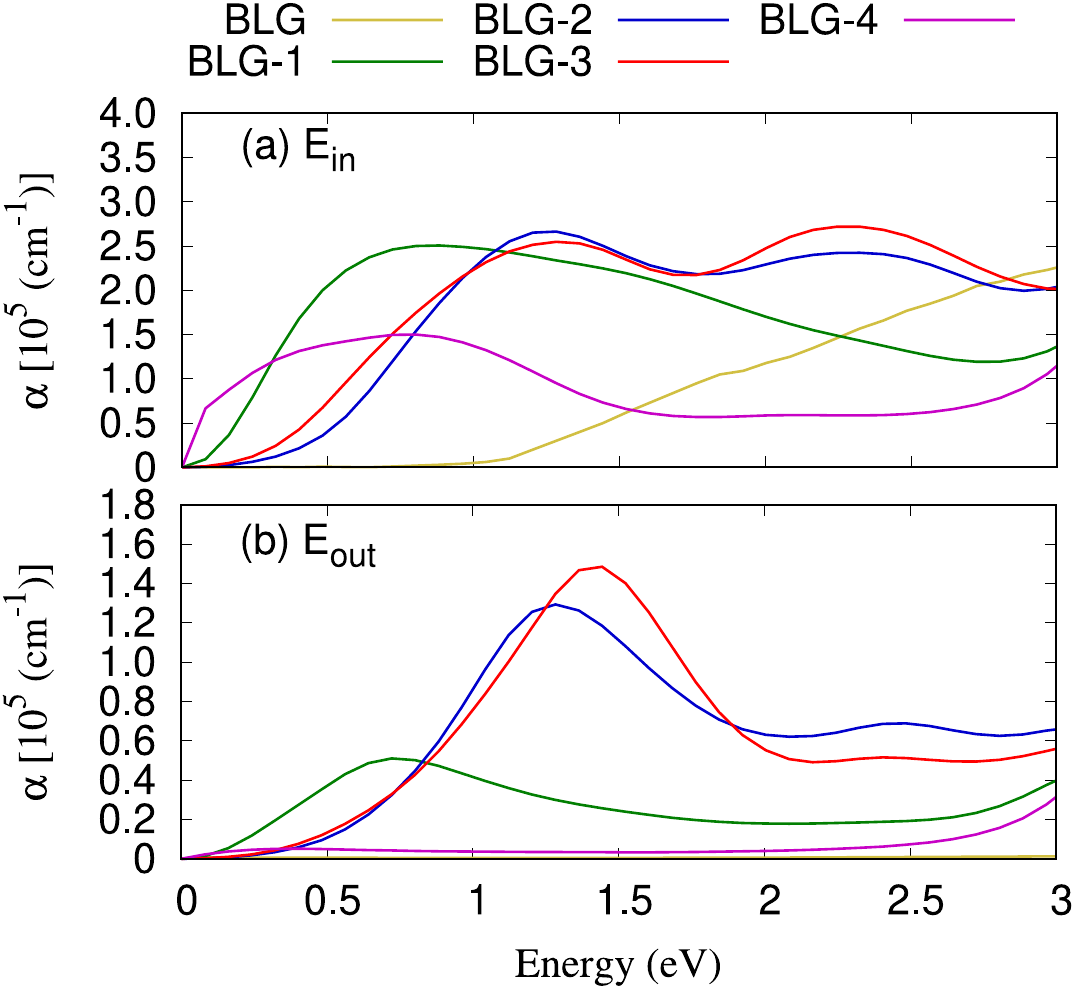}
	\caption{Absorption coefficient versus energy for pure BLG (golden),  BLG-1 (green), BLG-2  (blue), BLG-3 (red), and BLG-4 (purple),
		in the E$_{\rm in}$ (a), and  E$_{\rm out}$ (b).}
	\label{fig07}
\end{figure}

In our pure BLG, two main peaks in the imaginary part of the dielectric function are formed at $4.01$ and $14.03$~eV representing $\pi$ to $\pi^*$ transitions when the applied electric field is parallel to the BLG structure, and two peaks at $11.4$ and $14.42$~eV corresponding to the $\sigma$ and $\sigma^*$ transition are seen for perpendicular applied electric field (not shown). We observe a red shift in all four peaks in the presence of BN-codoped BLG with more or less the same intensity of the peaks (not shown). The red shift of the peaks is caused by the decreased energy spacing between
the $\pi$ and the $\pi^*$, and the $\sigma$ and the $\sigma^*$ along the $\Gamma$-M and the M-K directions (see \fig{fig02}). The detailes of these transitions are shown in our recent publication \cite{ABDULLAH2020100740}. In addition to these peaks, extra peaks in the low energy range from $0$ 
to $3.0$~eV, towards the visible range of radiation, are found corresponding to the bandgap opening 
of the structures. 
 
In this work we focus on the peaks formed at the low energies.
In \fig{fig05} and \fig{fig06} the imaginary part of dielectric function and the absorption coefficient are shown, respectively, for a parallel or in-plane, E$_{\rm in}$, (a) and a perpendicular or out of plane, E$_{\rm out}$, (b) electric field applied to the BLG structure \cite{NATH2015691, doi:10.1063/1.4904907, Nzar_ChinesePhysicsB2016}. 
The peaks in the imaginary part of the dielectric function in the presence of E$_{\rm in}$ are due to optical transitions in the bandgap or very close to the bandgap. We therefore see peaks formed at different values of energy indicating the bandgap energy. We should remember that values of the bandgaps are underestimate as the DFT with a local density approximation has been used here. Therefore, the peak position is not exactly equal to the bandgap energy.
Furthermore, the peaks in the presence of E$_{\rm out}$ represent the $\sigma$ and $\sigma^*$ transitions around the $\Gamma$-points.
It can clearly be seen that the peak intensity of $\varepsilon_2$ in the presence of E$_{\rm out}$ for BLG-2 and BLG-3 are maximum while a minimum peak intensity is found for BLG-4.

\begin{figure}[htb]
	\centering
	\includegraphics[width=0.4\textwidth]{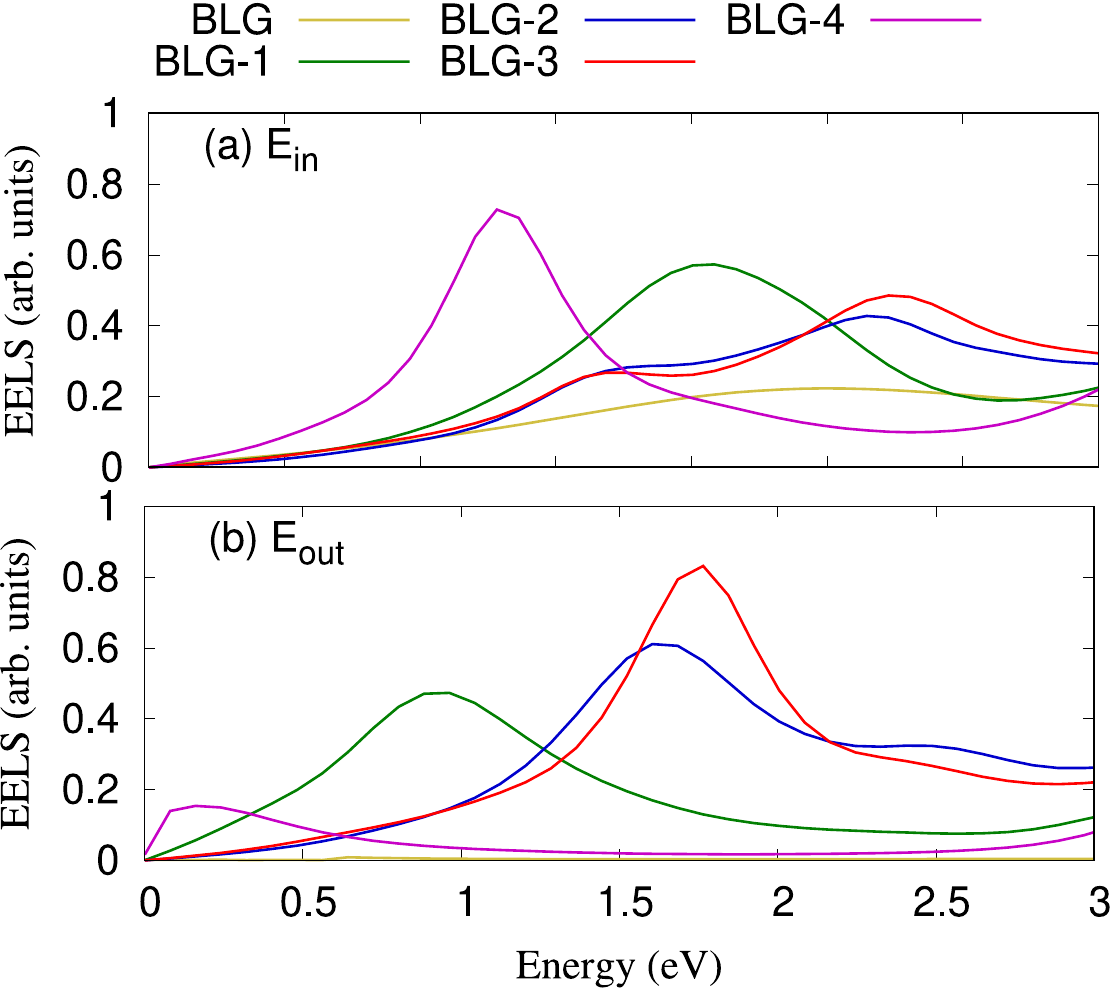}
	\caption{Energy loss-function, EELS, versus light energy for pure BLG (golden), BLG-1 (green), BLG-2 (blue), BLG-3 (red), and BLG-4 (purple), in the E$_{\rm in}$ (a), and  E$_{\rm out}$ (b).}
	\label{fig08}
\end{figure}

In addition, the energy loss-function, EELS characterizing inelastic scattering processes of the BN-codoped BLG structures is calculated for light polarization E$_{\rm in}$ and E$_{\rm out}$, and compared to BLG. The EELS is displayed in \fig{fig08} for E$_{\rm in}$ (a), and E$_{\rm out}$ (b)
in a low energy range.

We observe from the plots of EELS that with increasing the bandgap of the BN-codoped structures the 
peak is shifted to higher energy indicating 
that an inelastic scattering process occurs at a high energy of the light for the structure with a larger bandgap.

\section{Conclusion}\label{Sec:Conclusion}

This study uses DFT techniques based on the local density (LDA) approach to investigate electronic, mechanical, thermal and optical properties of BLG systems with Boron and Nitrogen dopant atoms. The DFT is implemented in Quantum espresso code and the Boltztrap package is used for further properties of the system. We show that the bandgap of AA-stacked BN-codoped BLG decreases with increasing interlayer spacing, while the bandgap of AB-stacked BN-codoped BLG increases. In addition, the attractive interaction between the B and the N atoms deteriorates the mechanical, thermal and optical properties of the system. In contrast, a repulsive interaction improve the high mechanical, thermal 
and optical characteristics of the system. Our results are relevant for optoelectronic applications of graphene-based devices. 

\section{Acknowledgment}
This work was financially supported by the University of Sulaimani and 
the Research center of Komar University of Science and Technology. 
The computations were performed on resources provided by the Division of Computational 
Nanoscience at the University of Sulaimani.  
 


\end{document}